\documentclass[twocolumn,twocolappendix]{aastex7}

\usepackage{amsmath}

\usepackage{booktabs}   
\usepackage{multirow} 
\shorttitle{Deep constraints on afterglow emission from AT2025ulz}
\shortauthors{O'Connor et al.}
\submitjournal{ApJL}

\usepackage{siunitx}
\usepackage{booktabs}
\usepackage{longtable}
\usepackage{caption}

\definecolor{blazeorange}{rgb}{1.0, 0.4, 0.0}
\definecolor{seagreen}{rgb}{0.18, 0.55, 0.34}
\definecolor{darkgreen}{rgb}{0.08, 0.45, 0.2}
\definecolor{rufous}{rgb}{0.66, 0.11, 0.03}
\definecolor{royalfuchsia}{rgb}{0.79, 0.17, 0.57}
\definecolor{scarlet}{rgb}{1.0, 0.13, 0.0}
\definecolor{royalpurple}{rgb}{0.47, 0.32, 0.66}

\begin{document}

\title{
AT2025ulz and S250818k: Deep X-ray and radio limits on off-axis afterglow emission and prospects for future discovery
}

\correspondingauthor{Brendan O'Connor}
\author[0000-0002-9700-0036]{Brendan O'Connor}
    \altaffiliation{McWilliams Fellow}
    \affiliation{McWilliams Center for Cosmology and Astrophysics, Department of Physics, Carnegie Mellon University, Pittsburgh, PA 15213, USA}
    \email[show]{boconno2@andrew.cmu.edu}

\author[0000-0003-4631-1528]{Roberto Ricci}
\affiliation{Department of Physics, University of Rome ``Tor Vergata'', via della Ricerca Scientifica 1, I-00133 Rome, Italy}
\affiliation{INAF-Istituto di Radioastronomia, Via Gobetti 101, I-40129 Bologna, Italy}
\email{ricci@ira.inaf.it}

\author[0000-0002-1869-7817]{Eleonora Troja}
\affiliation{Department of Physics, University of Rome ``Tor Vergata'', via della Ricerca Scientifica 1, I-00133 Rome, Italy}
\email{eleonora.troja@uniroma2.it}

\author[0000-0002-6011-0530]{Antonella Palmese}
\affiliation{McWilliams Center for Cosmology and Astrophysics, Department of Physics, Carnegie Mellon University, Pittsburgh, PA 15213, USA}
\email{palmese@cmu.edu}

\author[0000-0003-0691-6688]{Yu-Han Yang}
    \affiliation{Dipartimento di Fisica, Universit\`a di Tor Vergata, Via della Ricerca Scientifica, 1, 00133 Rome, Italy}
    \email{yyang@roma2.infn.it}

\author[0000-0001-9068-7157]{Geoffrey Ryan}
\affiliation{Perimeter Institute for Theoretical Physics, Waterloo, Ontario N2L 2Y5, Canada}
\email{gryan@perimeterinstitute.ca}

\author[0000-0002-8680-8718]{Hendrik van Eerten}
\affiliation{Department of Physics, University of Bath, Building 3 West, Bath BA2 7AY, United Kingdom}
\email{hjve20@bath.ac.uk }

\author[0009-0004-9520-5822]{Muskan Yadav}
\affiliation{Dipartimento di Fisica, Universit\`a di Tor Vergata, Via della Ricerca Scientifica, 1, 00133 Rome, Italy} 
\email{muskan.yadav@students.uniroma2.eu}

\author[0000-0002-9364-5419]{Xander J. Hall}
\affiliation{McWilliams Center for Cosmology and Astrophysics, Department of Physics, Carnegie Mellon University, Pittsburgh, PA 15213, USA}
\email{xhall@cmu.edu}

\author[0000-0003-3433-2698]{Ariel Amsellem}
\affiliation{McWilliams Center for Cosmology and Astrophysics, Department of Physics, Carnegie Mellon University, Pittsburgh, PA 15213, USA}
\email{aamselle@andrew.cmu.edu}

\author[0000-0002-0216-3415]{Rosa L. Becerra}
    \affiliation{Dipartimento di Fisica, Universit\`a di Tor Vergata, Via della Ricerca Scientifica, 1, 00133 Rome, Italy}
    \email{rosa.becerra@roma2.infn.it}

\author[0009-0001-0574-2332]{Malte Busmann}
    \affiliation{University Observatory, Faculty of Physics, Ludwig-Maximilians-Universität München, Scheinerstr. 1, 81679 Munich, Germany}
    \email{m.busmann@physik.lmu.de}

\author[0000-0002-1270-7666]{Tom\'as Cabrera}
\affiliation{McWilliams Center for Cosmology and Astrophysics, Department of Physics, Carnegie Mellon University, Pittsburgh, PA 15213, USA}
\email{tcabrera@andrew.cmu.edu}

\author[0000-0001-6849-1270]{Simone Dichiara}
    \affiliation{Department of Astronomy and Astrophysics, The Pennsylvania State University, 525 Davey Lab, University Park, PA 16802, USA}
    \email{sbd5667@psu.edu}

\author[0000-0001-7201-1938]{Lei Hu}
\affiliation{McWilliams Center for Cosmology and Astrophysics, Department of Physics, Carnegie Mellon University, Pittsburgh, PA 15213, USA}
\email{leihu@andrew.cmu.edu}

\author[0009-0005-1871-7856]{Ravjit Kaur}
\email{ravkaur@ucsc.edu}
\affiliation{Department of Astronomy and Astrophysics, University of California Santa Cruz, 1156 High St, Santa Cruz, CA 95064, USA}

\author[0009-0000-4830-1484]{Keerthi Kunnumkai}
\affiliation{McWilliams Center for Cosmology and Astrophysics, Department of Physics, Carnegie Mellon University, Pittsburgh, PA 15213, USA}
\email{kkunnumk@andrew.cmu.edu}

\author[0000-0003-2362-0459]{Ignacio Maga\~na~Hernandez}
\altaffiliation{McWilliams Fellow}
\affiliation{McWilliams Center for Cosmology and Astrophysics, Department of Physics, Carnegie Mellon University, Pittsburgh, PA 15213, USA}
\email{imhernan@andrew.cmu.edu}

\begin{abstract}

The first joint electromagentic (EM) and gravitational wave (GW) detection, known as GW170817, marked a critical juncture in our collective understanding of compact object mergers. However, it has now been 8 years since this discovery, and the search for a second EM-GW detection has yielded no robust discoveries. Recently, on August 18, 2025, the LIGO-Virgo-KAGRA collaboration reported a low-significance (high false alarm rate) binary neutron star merger candidate S250818k. Rapid optical follow-up revealed a single optical candidate AT2025ulz ($z$\,$=$\,$0.08484$) that initially appeared consistent with kilonova emission. We quickly initiated a set of observations with \textit{Swift}, \textit{XMM-Newton}, \textit{Chandra}, and the Very Large Array to search for non-thermal afterglow emission. 
Our deep X-ray and radio search rules out that the optical rebrightening of AT2025ulz is related to the afterglow onset, reinforcing its classification as a stripped-envelope supernova (SN 2025ulz). We derive constraints on the afterglow parameters for a hypothetical binary neutron star merger at the distance of AT2025ulz ($\approx$\,400 Mpc) based on our X-ray and radio limits. We conclude that our observational campaign could exclude a GW170817-like afterglow out to viewing angles of $\theta_\textrm{v}$\,$\approx$\,$12.5$ degrees. We briefly discuss the prospects for the future discovery of off-axis afterglows. 

\end{abstract}



\section{Introduction} 
\label{sec:intro}

The first binary neutron star (BNS) merger GW170817 \citep{Abbott+17-GW170817A-MMO} was a watershed moment in multimessenger astronomy. 
GW170817 and its electromagnetic counterparts provided unprecedented insight into the outflows launched during and after a BNS merger. For the first time, GW170817 showed that the jets of gamma-ray bursts have a complex angular structure through the discovery of off-axis afterglow emission \citep{Troja2017,Troja2018a,Troja2019a,Troja2020,Troja2022a,Hallinan2017,Lamb2017jet,Lazzati2018,Resmi2018,Mooley2018,Mooley2022,D'Avanzo2018,Xie2018,Margutti2018,GG2018,Ghirlanda2019,Ghirlanda2022,Ryan2024}. The detection of non-thermal emission following the discovery of a future kilonova candidate \citep[see, e.g.,][]{Kaur} can provide a robust determination of the source as a BNS merger and provide additional constraints on both its merger and jet properties. 

Recently, a kilonova candidate AT2025ulz \citep{2025GCN.41414....1S,2025GCN.41433....1H,2025GCN.41452....1O} was identified both spatially and temporally coincident with a low-significance sub-solar binary neutron star merger candidate S250818k \citep{2025GCN.41437....1L}. While the first few days of data were potentially consistent with kilonova emission \citep{Hall2025sn,Yang2025ulz,Kasliwal2025sn,Gillanders25ulz,Franz2025}, the source quickly rose  after $\sim$\,$5$ days. 
The observed rise was potentially consistent with the onset of off-axis afterglow emission, reminiscent of GW170817 \citep{Troja2017,Ryan2024}, 
thus motivating a deep search for non-thermal radiation with X-ray and radio observations.
Spectroscopic features characteristic of a Type IIb supernova, denoted SN 2025ulz, were later identified \citep{2025GCN.41532....1B,Hall2025desi,Gillanders25ulz,Kasliwal2025sn,Franz2025}. Prior to the identification of robust spectroscopic features, the flattening and subsequent rise of the source was potentially consistent with the onset of off-axis afterglow emission \citep{Kasliwal2025sn,Hall2025sn}. 

In this work, we report on our extensive campaign of multi-wavelength observations spanning X-ray to radio wavelengths. We present our constraints on off-axis afterglow emission from AT2025ulz, under the hypothetical condition that it was indeed a BNS merger, and discuss the prospects for future detection of off-axis afterglows from BNS mergers. The paper is laid out as follows. In \S \ref{sec:obs}, we present our X-ray and radio observations. Our afterglow constraints are outlined in \S \ref{sec:results}, and a discussion of the results is presented in \S \ref{sec:discuss}. We state our conclusions in \S \ref{sec:conclusions}. 


Throughout the manuscript we adopt a standard $\Lambda$CDM cosmology \citep{Planck2020} with $H_0$\,$=$\,$67.4$ km s$^{-1}$ Mpc$^{-1}$, $\Omega_\textrm{m}$\,$=$\,$0.315$, and $\Omega_\Lambda$\,$=$\,$0.685$. At the redshift $z$\,$=$\,$0.08484$ of AT2025ulz \citep{Hall2025desi}, we adopt the distance of 401 Mpc. All upper limits are reported at the $3\sigma$ level.

\section{Observations}
\label{sec:obs}

\subsection{Neil Gehrels Swift Observatory}
\label{sec:Swift}

The \textit{Neil Gehrels Swift Observatory} X-ray Telescope observed AT2025ulz on 2025-08-19 (PI: R. Stein) and 2025-08-22  (PI: R. Becerra; Table \ref{tab:xray_obs}), but did not detect any source \citep{2025GCN.41453....1H,2025GCN.41528....1B}. We used the Living Swift XRT Point Source Catalogue \citep[LSXPS;][]{LSXPS} upper limit server\footnote{\url{https://www.swift.ac.uk/LSXPS/ulserv.php}} to derive $3\sigma$ upper limits of $<2.1\times10^{-3}$ and $<6.1\times10^{-3}$ cts s$^{-1}$, respectively. We adopt a typical powerlaw spectrum for GRB afterglows assuming that the slope $p$ of the electron's powerlaw energy distribution $N(E)$\,$\propto$\,$E^{-p}$ is $p$\,$=$\,$2.2$ which leads to a photon index of $\Gamma$\,$=$\,$1.6$ for emission between the peak frequency and the cooling frequency \citep{Granot2002}. We note that $p$\,$=$\,$2.2$ is typical of the values derived from particle acceleration simulations \citep{Sironi2015}, and is consistent with the value inferred for GW170817 of $\Gamma$\,$=$\,$1.585$ \citep{Troja2019a}, which leads to $p$\,$\approx$\,$2.17$ \citep{Margutti2018,Troja2019a}. 
Additionally, X-ray emission from short GRBs generally lie in this regime between the peak frequency and the cooling frequency \citep[e.g.,][]{Fong2015,OConnor2020}.
Therefore, assuming photon index $\Gamma$\,$=$\,$1.6$ and Galactic hydrogen column density $N_\textrm{H}$\,$=$\,$2.5\times10^{20}$ cm$^{-2}$ \citep{Willingale2013}, we derive $0.3$\,$-$\,$10$ keV upper limits of $<8.9\times10^{-14}$ and $<2.6\times10^{-13}$ erg cm$^{-2}$ s$^{-1}$, respectively, to the unabsorbed flux. 

\subsection{XMM-Newton}
\label{sec:XMM}


We performed a Target of Opportunity (ToO) observation of AT2025ulz with \textit{XMM-Newton} starting on 2025-08-26 03:01:19 for 54 ks (PI: Troja; ObsID: 0964050101), see Table \ref{tab:xray_obs}. 
The three detectors (pn, MOS1 and MOS2) on the European Photon Imaging Camera (EPIC; \citealt{Struder2001,Turner2001}) were operated in full window mode, with the thin optical-blocking filter. The data were reduced with the Science Analysis System (\texttt{SAS}; \citealt{2004ASPC..314..759G}) v22.1 using the most recent calibration files. The exposure times in good time intervals (i.e. excluding high-rate flaring particle background), are 34.2 ks, 42.3 ks and 27.3 ks for detectors MOS1, MOS2, and pn, respectively.
Source counts were extracted from a circular region with a radius of $20\arcsec$, centered on the coordinates of AT2025ulz. The background was estimated from a source-free annulus with radii of $50\arcsec$\,$-$\,$100\arcsec$, centered on the same position (excluding one contaminating source) for both MOS detectors, and from a nearby circular region with a radius of $40\arcsec$ for the pn detector. The background counts were then rescaled to the source aperture using the ratio of the extraction areas.

At the location of AT2025ulz we do not detect any source.  We derive the $3\sigma$ upper limit to the count rate following \citet{Kraft1991}. We identify 37 total counts in the source region with 34 expected background counts in MOS1, 27 total counts with 37 background counts in MOS2, and 69 total counts with 77 background counts in pn. 
This yields upper limits of $<$\,$1.0\times10^{-3}$ cts s$^{-1}$, $<$\,$4.8\times10^{-4}$ cts s$^{-1}$, and $<$\,$1.2\times10^{-3}$ cts s$^{-1}$ for MOS1, MOS2, and pn, respectively. 
We derive a combined upper limit to the count rate of $<$\,$4.0\times10^{-4}$ cts s$^{-1}$. We have corrected for the encircled energy fraction. 

The generated spectral files were fit with  \texttt{XSPEC v12.14} \citep{Arnaud1996} using an absorbed power law model (using $\Gamma$\,$=$\,$1.6$ and $N_\textrm{H}$\,$=$\,$2.5\times10^{20}$ cm$^{-2}$; see \S \ref{sec:Swift}) to determine the energy correction factor for MOS1, MOS2, and pn. 
Putting this all together, we obtain an unabsorbed flux upper limit ($3\sigma$) of $<2.5\times10^{-15}$ erg cm$^{-2}$ s$^{-1}$ in the $0.3$\,$-$\,$10$ keV bandpass. 


\subsection{Chandra X-ray Observatory}
\label{sec:chandra}








We obtained X-ray observations with the \textit{Chandra X-ray Observatory} (CXO) of AT2025ulz through program 26400095\footnote{\url{https://doi.org/10.25574/cdc.488}} (PI: O'Connor). Our data was obtained with ACIS-S across two epochs (Table \ref{tab:xray_obs}) starting at 19.41 d for 49.41 ks and 42.87 d for 47.59 ks. The \textit{Chandra} data were retrieved from the \textit{Chandra} Data Archive (CDA)\footnote{\url{https://cda.harvard.edu/chaser/}} and processed using the \texttt{CIAO v4.17.0} data reduction package \citep{Ciao} with \texttt{CALDB v4.11.6}. At the position of AT2025ulz we do not detect an X-ray source in either epoch. In the first epoch, we identify 0 photons within a circular aperture of radius $1.0\arcsec$ and derive a count rate upper limit of $<9.4\times10^{-5}$ cts s$^{-1}$. This rate is corrected for the encircled energy fraction accounting for the \textit{Chandra} ACIS-S point-spread function. In the second epoch, we identify 1 photon and derive a count rate of $<1.8\times10^{-4}$ cts s$^{-1}$. We adopt a typical spectral shape of an absorbed powerlaw (using $\Gamma$\,$=$\,$1.6$ and $N_\textrm{H}$\,$=$\,$2.5\times10^{20}$ cm$^{-2}$; see \S \ref{sec:Swift}). This yields upper limits of $<2.1\times10^{-15}$ and $<4.2\times10^{-15}$ erg cm$^{-2}$ s$^{-1}$, respectively, to the $0.3$\,$-$\,$10$ keV unabsorbed flux.

\subsection{Very Large Array}
\label{sec:VLA}


We observed AT2025ulz with the Karl G. Jansky Very Large Array (VLA) starting on 2025-08-21 in X-band centered at 10 GHz with a bandwidth of 4 GHz \citep{Ricci-1} and on 2025-08-24 in S- and C-band at the center frequencies of 3 and 6 GHz, respectively \citep{Ricci-2}, under the program 22B-275 (PI: Troja). The array configuration was moving from C to B at the time of these observations. We obtained additional observations under joint \textit{Chandra}-VLA program SC260095 (PI: O'Connor) on 2025-09-08 and 2025-09-29 in C-band at a center frequency of 6 GHz with a bandwidth of 4 GHz for 2 hours of total observing time per run. In all the observations the primary calibrator was 3C286 and the phase calibrator was J1602+3326. The data were downloaded from the NRAO archive and calibrated in \texttt{CASA} \citep{CASA2022} using the VLA pipeline v6.6.1. The imaging was performed in \texttt{CASA v.5.5.0} using the task {\tt tclean} with a Briggs parameter value of 0.5 and H\"ogbom cleaning mode. As the array configuration was changing, for reference, we report that in our S-band observation the beam size is $1.64\arcsec\times1.56\arcsec$ with position angle (PA) of 27.9 deg and in our final C-band observation the beam was $1.03\arcsec\times0.95\arcsec$ with a PA of 53.8 deg. 
The final processed images were inspected with the \texttt{CASA} viewer and the root mean square (rms) noise per beam was computed in a region of the radio map away from bright sources using the \texttt{CASA} task {\tt imstat}. We do not detect a source at the location of AT2025ulz in any of our images. The resulting $3\sigma$ upper limits are tabulated in Table \ref{tab:radio_obs}.  




We note that our results are not in disagreement with the reported MeerKAT detection of radio emission at 3 GHz \citep[$\sim$\,$80$\,$-$\,$90\mu$Jy;][]{Bruni1,Bruni3,Rhodes1}, since the observations with the two instruments were characterized by different beam sizes. Whereas our S-band observations rule out a point-source to $<47\mu$Jy per beam \citep{Ricci-2}, the larger MeerKAT beam is resolving diffuse emission, likely due to star formation within the galaxy.


\begin{figure}
    \centering
    \includegraphics[width=\linewidth]{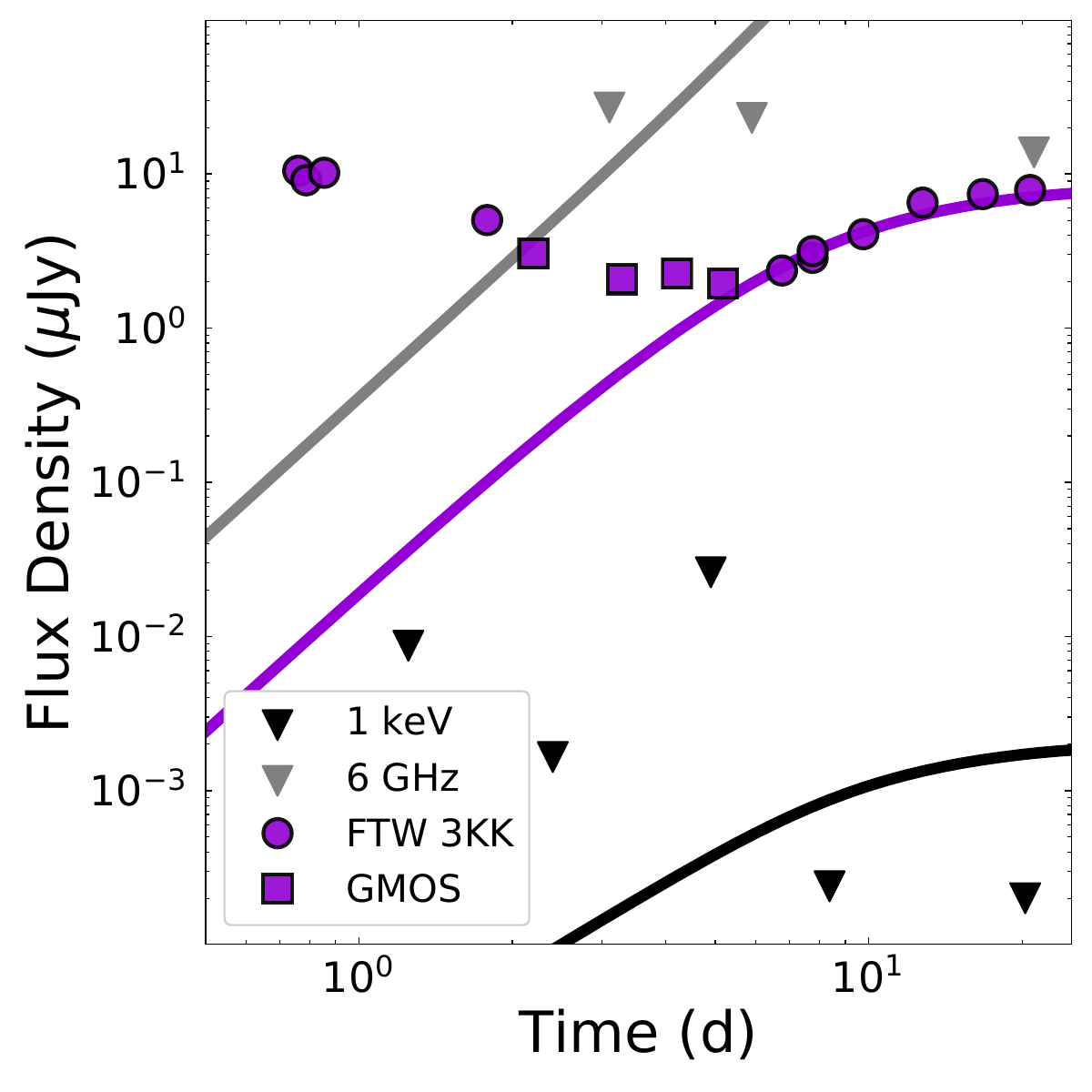}
    \caption{Optical lightcurve ($g$-band) of AT2025ulz out to 25 days from discovery compared to an off-axis afterglow model. Optical data from FTW 3KK and Gemini GMOS are reproduced from \citet{Hall2025sn}. X-ray (black) and radio (gray) upper limits from this work are shown (Tables \ref{tab:xray_obs} and \ref{tab:radio_obs}) and rule out that the optical rebrightening is due to an off-axis afterglow.
    }
    \label{fig:lcrise}
\end{figure}

\begin{figure}
    \centering
    \includegraphics[width=\linewidth]{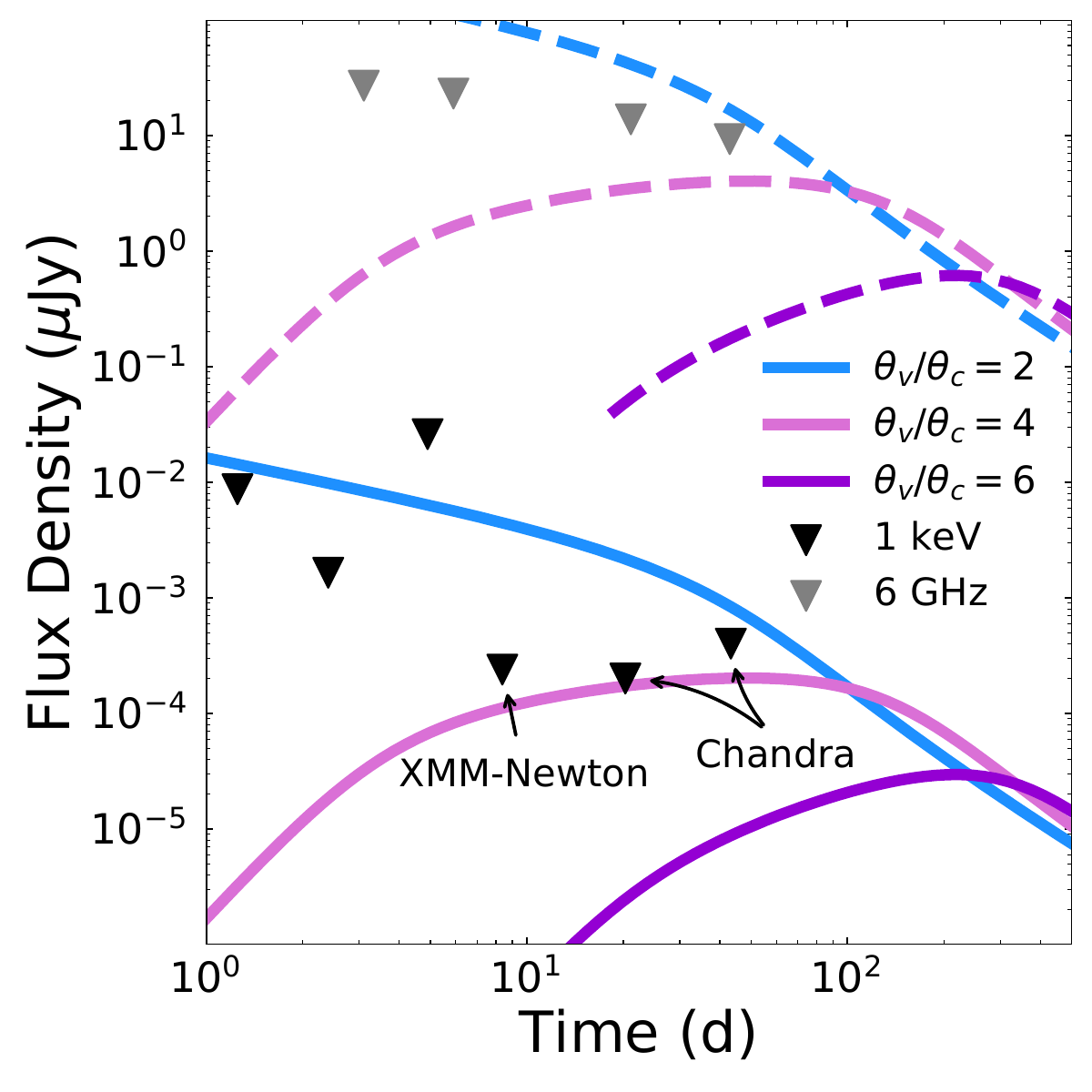}
    \caption{Afterglow lightcurves for different viewing angles $\theta_\textrm{v}/\theta_\textrm{c}$ generated using \texttt{afterglowpy} \citep{Ryan2020} using the afterglow fit parameters derived for GW170817 \citep{Ryan2024} for a distance of 400 Mpc. The viewing angle of $\theta_\textrm{v}/\theta_\textrm{c}$\,$=$\,$6$ corresponds to the observed value for GW170817. X-ray lightcurves at 1 keV are shown as solid lines and radio lightcurves at 6 GHz as dashed lines. X-ray upper limits are shown as black triangles and radio upper limits as gray triangles. All radio observations are from the VLA, and X-ray observations are ordered from left to right as: \textit{Swift}, EP/FXT, \textit{Swift}, \textit{XMM-Newton}, \textit{Chandra}, and \textit{Chandra}. The parameters used to plot the lightcurves are \citep{Ryan2024}: $E_\textrm{kin} = 4.8\times 10^{53}$ erg, $\theta_\textrm{c} = 3.2^\circ$, $\theta_\textrm{w} = 4.9 \theta_\textrm{c}$, $n = 2.4 \times 10^{-3} \mathrm{cm}^{-3}$, $p = 2.13$, $\varepsilon_\textrm{e} = 1.9\times 10^{-3}$, $\varepsilon_\textrm{B} = 5.75\times 10^{-4}$, $\xi_\textrm{N} = 1.0$, $d_\textrm{L} = 400$ Mpc, and $z = 0.08484$. 
    }
    \label{fig:lclimits}
\end{figure}

\section{Results}
\label{sec:results}

\subsection{A Rising Optical Lightcurve}

While AT2025ulz initially displayed a fast fading optical lightcurve (see Figure \ref{fig:lcrise}) that was plausibly consistent with kilonova emission \citep{Hall2025sn,Yang2025ulz,Kasliwal2025sn,Gillanders25ulz}, its lightcurve sharply rose after $\sim$\,$5$ days. In Figure \ref{fig:lcrise}, we show that this optical rise was potentially consistent with the off-axis afterglow produced by a relativistic jet \citep[see also][]{Kasliwal2025sn,Hall2025sn}. This motivated the need for deep X-ray and radio observations of the source. However, our deep  X-ray constraints (Table \ref{tab:xray_obs}) obtained with \textit{XMM-Newton} and \textit{Chandra} robustly rule out this possibility as it would require instead very significant X-ray detections near peak. The same is true for our radio observations (Table \ref{tab:radio_obs}), which are also capable of excluding this scenario. Additionally, optical spectroscopy identified features consistent with a Type IIb supernova \citep{2025GCN.41532....1B,Kasliwal2025sn,Gillanders25ulz}, firmly ruling out this possibility. This emphasizes the need for sensitive spectroscopic observations of future kilonova candidates, as well as deep X-ray and radio observations to provide a full picture of the source. In what follows, we expand upon the constraints available from our X-ray and radio observations.

\subsection{Comparison to GW170817}

\begin{figure}
    \centering
    \includegraphics[width=\linewidth]{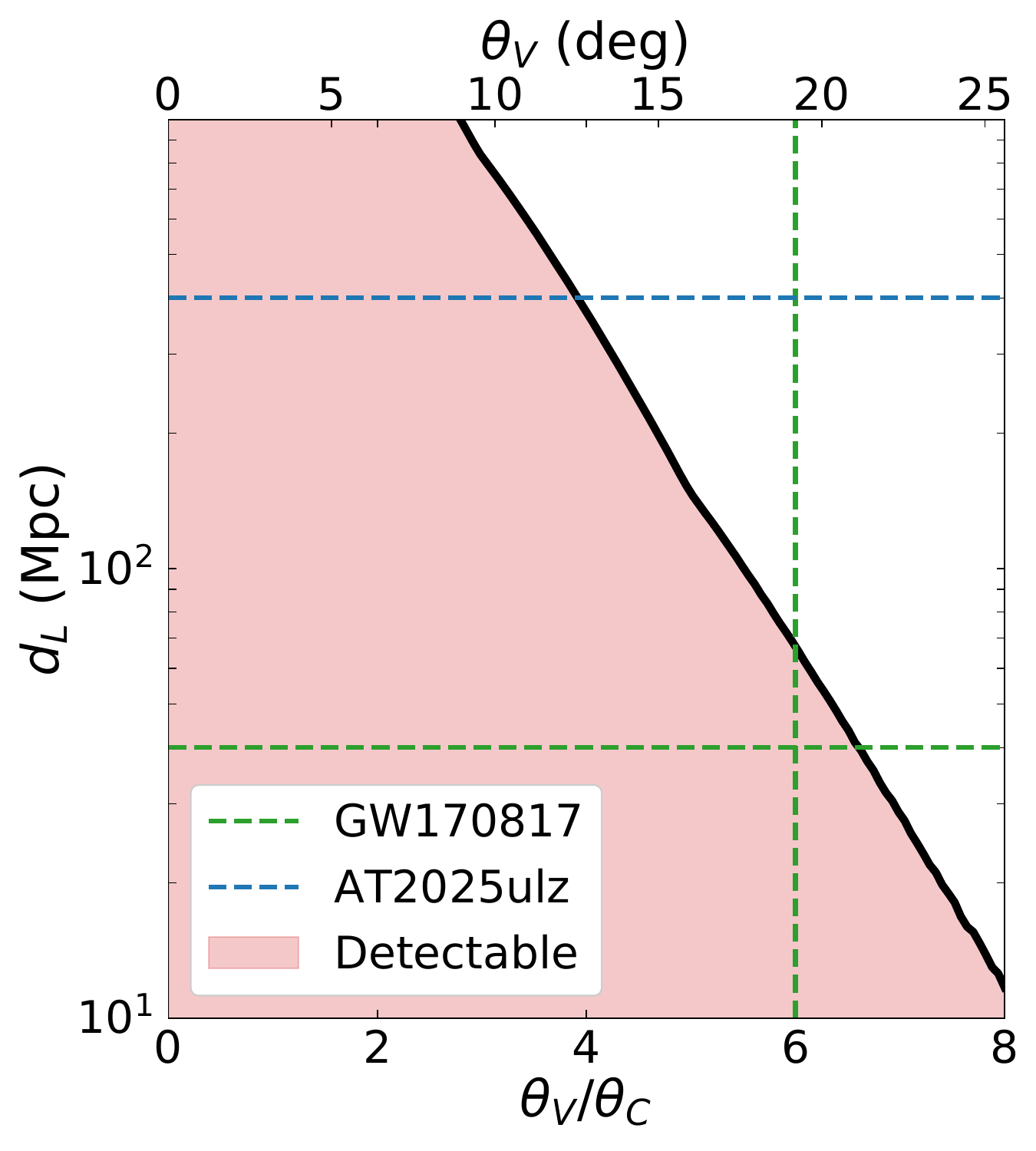}
    \caption{The maximum detectable distance of a GW170817-like afterglow \citep{Ryan2024} versus the viewing angle $\theta_\textrm{v}/\theta_\textrm{c}$ using our deep X-ray and radio limits. The distance of GW170817 (40 Mpc) and AT2025ulz (400 Mpc) are shown for reference as dashed horizontal lines. A dashed vertical line marks the viewing angle of GW170817 \citep{Ryan2024}. The afterglow parameters used to generate the lightcurves are the same as shown in the caption of Figure \ref{fig:lclimits}.
    }
    \label{fig:maxdet}
\end{figure}

In Figure \ref{fig:lclimits}, we compare our X-ray and radio upper limits to the off-axis afterglow of GW170817 \citep{Troja2017,Hallinan2017,Mooley2018,Mooley2022,Hotokezaka+19,Troja2020,Ghirlanda2022,Ryan2024,Palmese:2023beh}. We use limits from \textit{Chandra}, \textit{XMM-Newton}, and the VLA as well as the early X-ray upper limits from \textit{Swift}/XRT \citep{2025GCN.41453....1H,2025GCN.41528....1B} and EP/FXT \citep{2025GCN.41460....1L}. Despite the significantly larger distance (400 Mpc for AT2025ulz versus 40 Mpc for GW170817), we are still able to place robust constraints on a GW170817-like afterglow \citep{Ryan2024} out to viewing angles of $\theta_\textrm{v}/\theta_\textrm{c}$\,$\approx$\,$4$, where $\theta_\textrm{v}$ is the observer's inclination  with respect to the jet's axis and $\theta_\textrm{c}$ is the jet's core half-opening angle. As the viewing angle of GW170817 was $\theta_\textrm{v}/\theta_\textrm{c}$\,$\approx$\,$6$ \citep{Ryan2024}, our observations are not sensitive to such far off-axis jets at this distance. Therefore, we further compute the maximum detectable distance for GW170187 as a function of viewing angle $\theta_\textrm{v}/\theta_\textrm{c}$ in Figure \ref{fig:maxdet}. As GW170817 is only a single event in the broad diversity of short GRB afterglows \citep[e.g.,][]{Fong2015, Troja2019a}, we additionally compute detectability over a broad parameter space in \S \ref{sec:constraints}.

\subsection{Afterglow Parameter Constraints}
\label{sec:constraints}


Here we compute the allowed parameters for non-detection of a synchrotron afterglow \citep{Sari1998,Granot2002} using the \texttt{afterglowpy} package \citep{Ryan2020,Ryan2024}. We model the afterglow with a Gaussian structured jet propagating into a uniform density environment. The physical setup is specified by eight parameters: the isotropic-equivalent kinetic energy at the jet's core $E_\textrm{kin}$, the jet's core half-opening angle $\theta_\textrm{c}$, the observer's viewing angle $\theta_\textrm{v}$, the cutoff angle of the jet's structure $\theta_\textrm{w}$, the circumburst density $n$, the magnetic and electron energy fractions $\varepsilon_\textrm{B}$ and $\varepsilon_\textrm{e}$, and the electron power-law index $p$. We include an initial Lorentz factor at the jet's core of $\Gamma_0$\,$=$\,$300$, apply the same Gaussian angular profile for Lorentz factor $\Gamma(\theta_\textrm{v})$, and disable lateral jet spreading \citep[for details, see][]{Kaur}. We fixed the truncation angle $\theta_\textrm{w}$\,$=$\,$4.9\theta_\textrm{c}$ and have also fixed the electron participation fraction $\xi_\textrm{N}$\,$=$\,$1.0$. 

\begin{figure*}
    \centering
    \includegraphics[width=\linewidth]{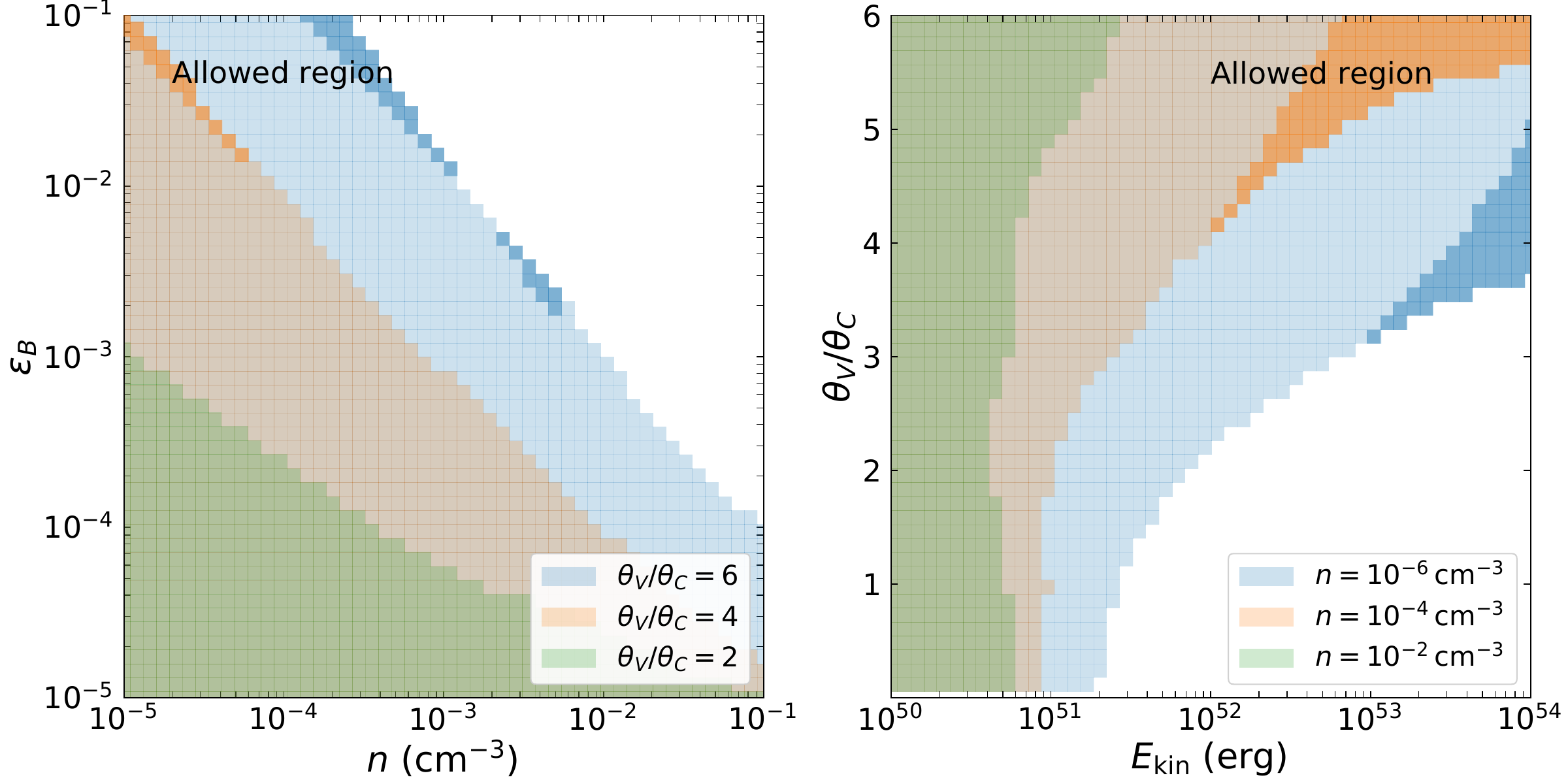}
    \caption{Allowed parameter space (shaded regions) for afterglow non-detection, assuming a Gaussian structured jet, at 400 Mpc ($z$\,$=$\,$0.08484$) based on our deep radio and X-ray upper limits. The darker shaded regions show a range of parameters that produce late-peaking afterglows that are detectable to our observational limits (if extended in time to $\sim$\,$150$ d), but not ruled out by the current data ($<$\,$50$ d). \textbf{Left:} Allowed values of the magnetic energy fraction $\varepsilon_\textrm{B}$ versus circumburst density $n$ for different viewing angles $\theta_\textrm{v}/\theta_\textrm{c}$. We have fixed $E_\textrm{kin}$\,$=$\,$10^{52}$ erg, $\theta_\textrm{c}$\,$=$\,$0.056$ rad (3.2 deg), $\varepsilon_\textrm{e}$\,$=$\,$0.1$, and $p$\,$=$\,$2.2$. \textbf{Right:} Allowed viewing angles $\theta_\textrm{v}/\theta_\textrm{c}$ as a function of core kinetic energy $E_\textrm{kin}$ for different circumburst density $n$. We have fixed $\theta_\textrm{c}$\,$=$\,$0.056$ rad (3.2 deg), $\varepsilon_\textrm{B}$\,$=$\,$10^{-2}$, $\varepsilon_\textrm{e}$\,$=$\,$0.1$, and $p$\,$=$\,$2.2$.
    }
    \label{fig:lc}
\end{figure*}

We generate afterglow models using these parameters and compare the flux density at each time and frequency to our X-ray and radio upper limits (see Tables \ref{tab:xray_obs} and \ref{tab:radio_obs}). If the flux density exceeds our limits at any time or frequency of our upper limits we consider those parameters excluded. As the reasonable range of afterglow parameter space is notably large, we focus on varying two parameters at a time holding all else fixed. A full search of the allowed parameter space is beyond this work, but we have explored a broad range of possible values for $n$, $\varepsilon_\textrm{B}$, $E_\textrm{kin}$, and viewing angle $\theta_\textrm{v}/\theta_\textrm{c}$. The results are shown in Figure \ref{fig:lc}. 

In Figures \ref{fig:maxdet} and \ref{fig:lc}, we have considered detectability based only on our observations (i.e., at specific times and frequencies; Tables \ref{tab:xray_obs} and \ref{tab:radio_obs}). However, this does not account for the possibility of late-peaking afterglows that may become detectable at $\gtrsim$\,$100$ day timescales. Therefore, we performed an additional check, considering the deepest limit at each frequency, and determine whether any lightcurves that are currently allowed by our observations could become detectable at later times. We find that there is a small slice of parameter space where this is possible, specifically requiring larger kinetic energies $\gtrsim$\,$10^{52}$ erg, small densities $\lesssim$\,$10^{-3}$ cm$^{-3}$, and large off-axis viewing angles $\theta_\textrm{v}/\theta_\textrm{c}$\,$\gtrsim$\,$3$, see the darker shaded regions in Figure \ref{fig:lc}. Each of these choices leads to later peaking afterglows which may become detectable at late-times even at the large distance (400 Mpc) of AT2025ulz. 

Motivated by this, we consider the impact of an additional epoch of \textit{Chandra} and VLA data at 150 days from discovery with the same depth. The region of parameter space that can be excluded with the addition of this single additional epoch is shown as a darker shaded region in Figure \ref{fig:lc}. We find that this is capable of excluding this small parameter space of late peaking afterglows that were not excluded by the current observations. For future BNS mergers, continued follow-up to late-times is strongly recommended to rule out late peaking lightcurves and provide the maximum constraints on the allowed parameter space.

\section{Discussion}
\label{sec:discuss}

\subsection{Future Prospects: Detectability of Off-axis Jets}

The detection of future off-axis afterglows at larger distances than GW170817 poses a major observational challenge \citep[e.g.,][]{Kaur}. Our comprehensive observational campaign of AT2025ulz provides a realistic case study of these challenges at 400 Mpc. While our observations are sensitive to GW170817 out to a viewing angle of $\theta_\textrm{v}/\theta_\textrm{c}$\,$\approx$\,$4$ (Figure \ref{fig:maxdet}), corresponding to $\sim$\,$12.5$ deg (compared to 20 deg for GW170817), they are not capable of detecting afterglows at the typical expected viewing angle of gravitational wave events ($\sim$\,$35$ deg; \citealt{Schutz2011}). While this sensitively depends on the assumed half-opening angle \citep[Figure \ref{fig:lc-2}; and see][]{Kaur}, the larger distances of future BNS mergers, compared to GW170817 at 40 Mpc, are difficult to reconcile. 

However, for different combinations of the jet microphysics $\varepsilon_\textrm{B}$ and density $n$, we are capable of probing these expected large off-axis angles even at 400 Mpc. The exact constraints depend slightly on the choice of core half-opening angle, and Figure \ref{fig:lc-2} shows these how the constraint changes for a $2\times$ larger core angle (i.e., $\theta_\textrm{c}$\,$\approx$\,$6$ deg). Notably, our understanding of short GRB jet opening angles is limited to around a dozen events and displays a broad range between $\sim$\,$2$ to $25$ deg  \citep[e.g.,][]{Burrows2006,Soderberg2006,Berger2013,Fong2012,Fong2014,Fong2015,Fong2021kn,Troja2016jetbreak,Troja2019b,Lamb2019grb160821B,OConnor2021kn,RoucoEscorial2022}. We have adopted for reference the value ($\approx$\,$3$ deg; \citealt{Ryan2024}) inferred for GW170817 (Figure \ref{fig:lc}) and the median value ($\approx$\,$6$ deg; \citealt{RoucoEscorial2022}) inferred from the population of short GRBs with measured jet breaks (Figure \ref{fig:lc-2}). Additional late-time X-ray and radio observations of short GRBs can increase this population of events, which will improve our measurements of the distribution of jet angles and therefore refine our predictions for future detections \citep[see, e.g.,][]{Kaur}.

In any case, for on-axis (or close to on-axis) jets, e.g., $\theta_\textrm{v}/\theta_\textrm{c}$\,$<$\,$2$, we can constrain a broad region of the reasonable parameter space of energy, density, and jet microphysics (Figure \ref{fig:lc}). For typical short GRB parameters \citep{Fong2015,OConnor2020,OConnor2024}, we can strongly disfavor such close viewing angles ($\theta_\textrm{v}/\theta_\textrm{c}$\,$<$\,$2$). Even for larger off-axis angles ($\theta_\textrm{v}/\theta_\textrm{c}$\,$\approx$\,$6$), as seen for GW170817 \citep{Mooley2018,Ryan2024}, our deep observations are capable of ruling out dense environments at the upper end ($>$\,$10^{-2}$ cm$^{-3}$) of the short GRB density distribution \citep{Fong2015,OConnor2020}. 

We further show in Figure \ref{fig:lc-3} the impact of our assumption of the powerlaw index $p$ of the electron's energy distribution. As there are no detections of an afterglow that would provide a measure of this index, we have assumed the typical value derived from particle acceleration simulations of $p$\,$\approx$\,$2.2$ \citep[e.g.,][]{Sironi2015}. This is also similar to the value of $p$\,$\approx$\,$2.17$ derived for GW170817 \citep{Margutti2018,Troja2019a}. We find that varying the value of $p$, for example to $p$\,$=$\,$2.6$ (Figure \ref{fig:lc-3}), does not significantly modify our afterglow constraints. We note that both of these values of $p$ are in the typical range of observed values for short GRBs, which span between $p$\,$\approx$\,$2$\,$-$\,$3$ \citep[e.g.,][]{Fong2015}. 

While the broad constraints placed by our deep X-ray and radio non-detections do not rule out all reasonable combinations of afterglow and jet parameters, they do show that for favorable combinations of these parameters, even at 400 Mpc (and beyond), the detection of future off-axis afterglows is possible and can provide robust constraints on the jet's of future BNS mergers. In the future, these observational limits can be paired with additional information based on the detection (or lack of detection) of prompt gamma-ray emission and constraints on the observer's viewing angle (at the specific distance and location of the BNS; see \citealt{Chen2019}) derived from gravitational radiation. Even in the absence of afterglow detections, this is a promising pathway to provide constraints on the jet properties.

\subsection{Comparison to Type II Supernova X-ray and Radio Lightcurves}






As AT2025ulz has been robustly classified as a Type IIb stripped-envelope supernova \citep{Hall2025sn,Hall2025desi,Kasliwal2025sn,Yang2025ulz,Gillanders25ulz}, we briefly compare our upper limits to the X-ray and radio lightcurves of other Type IIb supernovae. At 400 Mpc, our data reached an X-ray luminosity ($0.3$\,$-$\,$10$ keV) of roughly $\sim$\,$10^{40}$ erg s$^{-1}$ and radio luminosity of $\sim$\,$(2-4)\times10^{27}$ erg s$^{-1}$ Hz$^{-1}$ at 6 GHz. We find that our limits are sensitive to the most luminous Type IIb supernovae, but cannot exclude the majority of the population. 

In particular, at radio wavelengths we could detect SN 2003bg \citep{Soderberg2006sn} or SN 2016bas \citep{Bietenholz2021}, but we would not be sensitive to SN 1993J \citep{Weiler2007}. At X-rays wavelengths, our data would likely barely be sensitive to SN 1993J \citep{Chandra2009} at 8.5 days with \textit{XMM-Newton}. As SN 1993J is the brightest Type IIb at X-ray wavelengths, we can not probe the majority of the population due to the significantly larger distance of AT2025ulz.

\section{Conclusions}
\label{sec:conclusions}

We present deep X-ray and radio observations of AT2025ulz using \textit{Swift}, \textit{Chandra}, \textit{XMM-Newton}, and the VLA and place constraints on off-axis afterglow emission. While AT2025ulz was eventually classified a Type IIb supernova \citep{Hall2025sn,Yang2025ulz,Kasliwal2025sn,Gillanders25ulz}, these observations serve as a useful case study for afterglow searches from future binary neutron star merger candidates uncovered by LIGO-Virgo-KAGRA in future observing runs (e.g., O5, A\#; McIver et al., in preparation), especially as a BNS horizon continues to increase. In the future, deep X-ray and radio observations with next-generation facilities will be critical to detect off-axis emission at $>$\,$200$ Mpc \citep{Kaur}. Specifically, \textit{AXIS} \citep{mushotzky2019advanced}, and \textit{NewAthena} \citep{NewAthena} at X-ray wavelengths and the Square Kilometer Array \citep[SKA;][]{dewdney2009square} and the next-generation VLA \citep[ngVLA;][]{butler2018sensitivity,butler2019ngvla,Corsi2019} at radio wavelengths will be critical to achieving this science in the 2030s and beyond. 

\begin{acknowledgments}
BO acknowledges  Mansi Kasliwal for useful discussions. 
BO is supported by the McWilliams Postdoctoral Fellowship in the McWilliams Center for Cosmology and Astrophysics at Carnegie Mellon University. Support for this work was provided by the National Aeronautics and Space Administration through Chandra Award Number GO5-26015X issued by the Chandra X-ray Center, which is operated by the Smithsonian Astrophysical Observatory for and on behalf of the National Aeronautics Space Administration under contract NAS8-03060. ET, RR, YYH, and MY are supported by the European Research Council through the Consolidator grant BHianca (grant agreement ID~101002761). AP, TC, LH, IMH are supported by NSF Grant No. 2308193.

This work used resources on the Vera Cluster at the Pittsburgh Supercomputing Center (PSC). Vera is a dedicated cluster for the McWilliams Center for Cosmology and Astrophysics at Carnegie Mellon University. We thank the PSC staff for their support of the Vera Cluster.

The scientific results reported in this article are based on observations made by the Chandra X-ray Observatory. This paper employs a list of Chandra datasets, obtained by the Chandra X-ray Observatory, contained in the Chandra Data Collection \dataset[doi:10.25574/cdc.488]{https://doi.org/10.25574/cdc.488}. This research has made use of software provided by the Chandra X-ray Center (CXC) in the application package CIAO. This research has made use of data and/or software provided by the High Energy Astrophysics Science Archive Research Center (HEASARC), which is a service of the Astrophysics Science Division at NASA/GSFC. Based on observations obtained with XMM-Newton, an ESA science mission with instruments and contributions directly funded by ESA Member States and NASA. The National Radio Astronomy Observatory is a facility of the National Science Foundation operated under cooperative agreement by Associated Universities, Inc. This work made use of data supplied by the UK \textit{Swift} Science Data Centre at the University of Leicester. This research has made use of the XRT Data Analysis Software (XRTDAS) developed under the responsibility of the ASI Science Data Center (ASDC), Italy. 
\end{acknowledgments}





%
\facilities{\textit{Neil Gehrels Swift Observatory}, \textit{Chandra X-ray Observatory}, \textit{XMM-Newton}, Karl G. Jansky Very Large Array
}

\software{\texttt{Astropy} \citep{2018AJ....156..123A,2022ApJ...935..167A}, \texttt{afterglowpy} \citep{Ryan2020},  \texttt{XSPEC} \citep{Arnaud1996}, \texttt{CIAO} \citep{Ciao}, \texttt{HEASoft} \citep{2014ascl.soft08004N}, \texttt{SAS} \citep{2004ASPC..314..759G}, \texttt{CASA} \citep{CASA2022}
}


\appendix

\section{Log of Observations}

Here we present the log of observations analyzed in this paper (see Tables \ref{tab:xray_obs} and \ref{tab:radio_obs}).

\begin{deluxetable}{lcccccc}
\tablecaption{Summary of our X-ray Observations. The upper limits are at $3\sigma$ and in the $0.3$\,$-$\,$10$ keV energy range. The upper limits derived from \textit{Chandra} are summed across the 3 observations obtained in each epoch.
\label{tab:xray_obs}}
\tablewidth{0pt}
\tabletypesize{\footnotesize}
\tablehead{
\colhead{Start Date (UT)} & \colhead{$T+T_0$ (d)} & \colhead{Facility} & \colhead{Instrument} & \colhead{Exposure (ks)} & \colhead{ObsID} & \colhead{Flux (erg cm$^{-2}$ s$^{-1}$)}
}
\startdata
2025-08-19 08:21:15 & 1.29 & \textit{Swift} & XRT & 3.59 & 7400151001 & $<8.9\times10^{-14}$ \\
2025-08-22 21:56:17 & 4.86 & \textit{Swift} & XRT & 2.88 & 7400151002 & $<2.6\times10^{-13}$\\
2025-08-26 03:01:19 & 8.07 & \textit{XMM-Newton} & pn/MOS1/MOS2 & 54.00 & 0964050101 & $<2.5\times10^{-15}$ \\ 
2025-09-06 11:06:46 & 19.41 & \textit{Chandra} & ACIS-S &  9.78 & 29745 & $<2.1\times10^{-15}$ \\
2025-09-07 03:56:25 & 20.11 & \textit{Chandra} & ACIS-S & 24.75 & 31484 & -- \\
2025-09-07 16:46:46 & 20.64 & \textit{Chandra} & ACIS-S & 14.88 & 31485 & -- \\
2025-09-29 22:15:21 & 42.87 & \textit{Chandra} & ACIS-S & 14.88 & 29746 & $<4.2\times10^{-15}$ \\
2025-09-30 07:44:51 & 43.27 & \textit{Chandra} & ACIS-S & 15.86 & 31902 & -- \\
2025-09-30 17:47:40 & 43.69 & \textit{Chandra} & ACIS-S & 16.85 & 31903 & -- \\
\enddata
\end{deluxetable}

\begin{deluxetable}{lccccccc}
\tablecaption{Summary of our radio observations obtained with the VLA.
\label{tab:radio_obs}}
\tablewidth{0pt}
\tabletypesize{\footnotesize}
\tablehead{
\colhead{Start Date (UT)}& \colhead{$T+T_0$ (d)} &  \colhead{Facility} & \colhead{Band (GHz)} & \colhead{Exposure (s)} & \colhead{Flux density ($\mu$Jy)} & \colhead{Configuration} & \colhead{Program}
}
\startdata
2025-08-21 02:57 & 3.07 & VLA & X (10)   &  $20\times 60$ & $<27$ & C $\rightarrow$ B & 22B-275\\
2025-08-24 01:32 & 6.01 & VLA & C (6)  & $15\times 60$  & $<23$ & C $\rightarrow$ B & 22B-275 \\
2025-08-24 01:32 & 6.01 & VLA &  S (3) & $15\times 60$ & $<47$ & C $\rightarrow$ B & 22B-275 \\
 2025-09-08 03:50  & 21.10 & VLA & C (6) & $75\times 60$ & $<14$ & B & SC260095 \\
 2025-09-29 23:30 & 42.92 & VLA & C (6) & $75\times 60$ & $<9.3$ & B & SC260095\\
\enddata
\end{deluxetable}

\section{Additional Afterglow Constraints}

\begin{figure*}
    \centering
    \includegraphics[width=\linewidth]{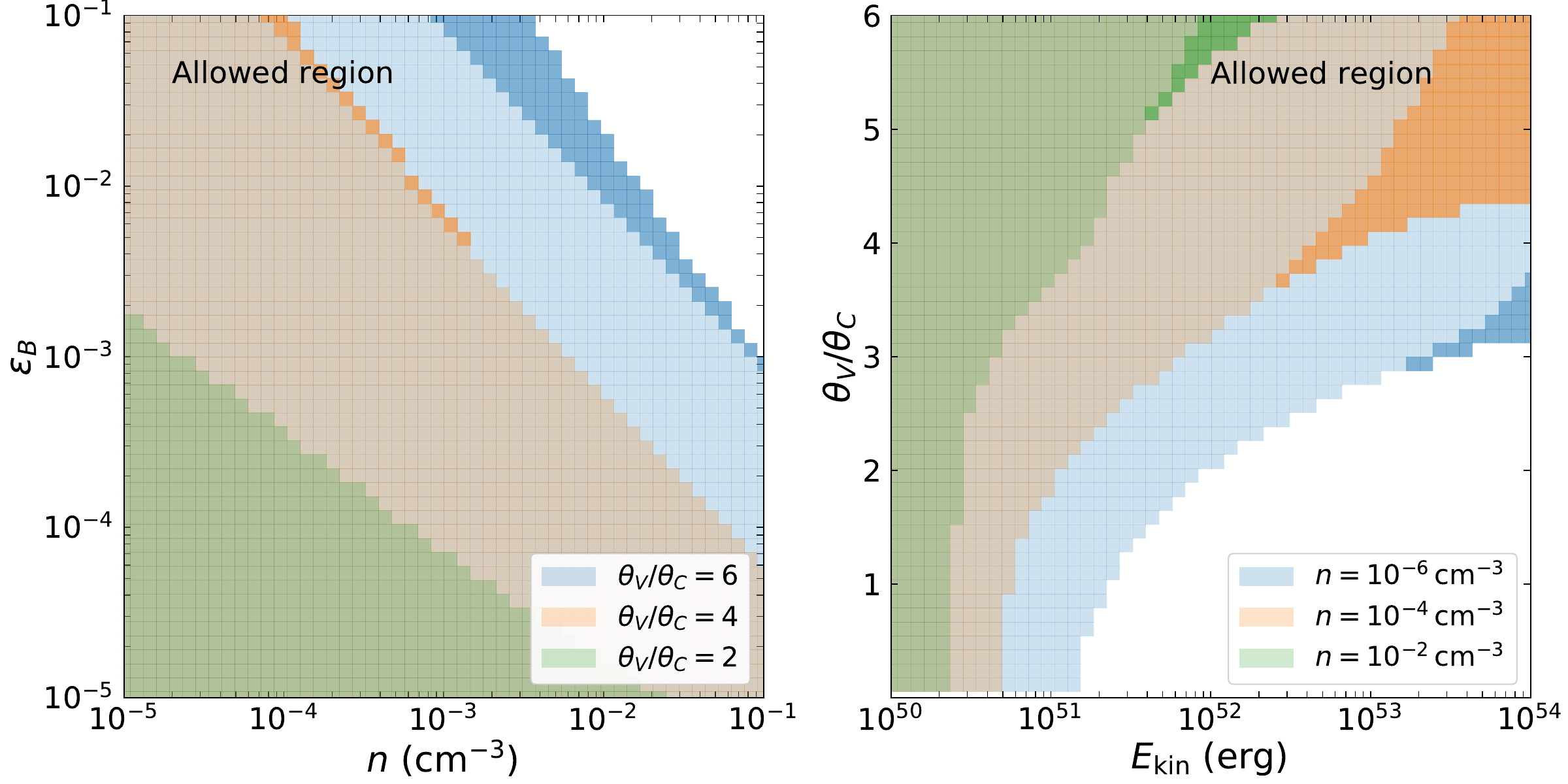}
    \caption{Same as Figure \ref{fig:lc} but for a wider jet $\theta_\textrm{c}$\,$=$\,$0.1$ rad. For wider jets we are less constraining on the other jet parameters. 
    }
    \label{fig:lc-2}
\end{figure*}

\begin{figure*}
    \centering
    \includegraphics[width=\linewidth]{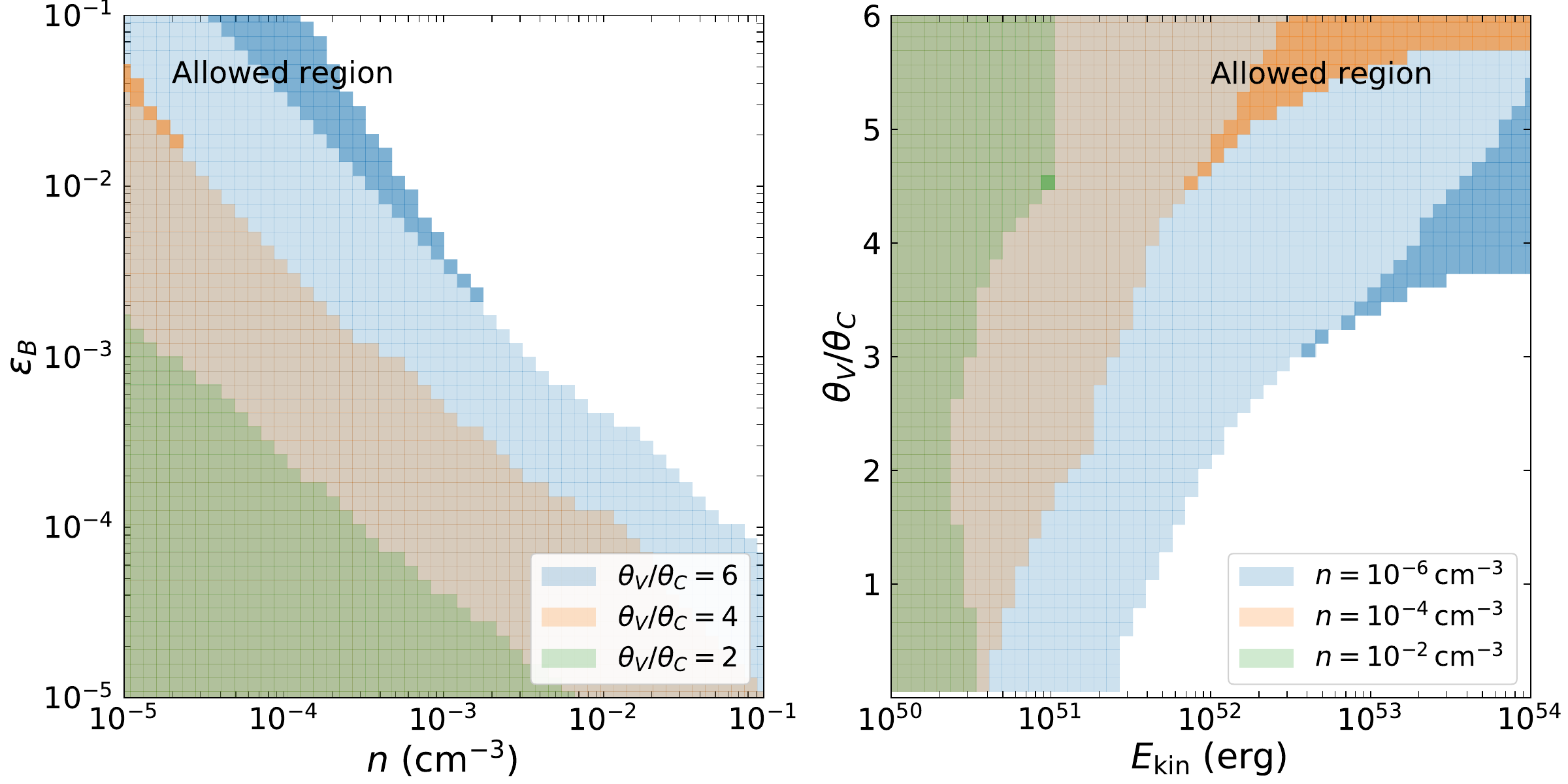}
    \caption{Same as Figure \ref{fig:lc} but for  $p$\,$=$\,$2.6$. 
    }
    \label{fig:lc-3}
\end{figure*}

Here we show additional afterglow parameter constraints in Figure \ref{fig:lc-2} and Figure \ref{fig:lc-3} to investigate the impact of changing $\theta_\textrm{c}$ and $p$, respectively. 

\bibliography{bib}{}
\bibliographystyle{aasjournal}



\end{document}